\documentclass[12pt,preprint]{aastex} 
\usepackage{color}
\usepackage{epsfig}
\usepackage{wrapfig}
\usepackage{bm}

\usepackage{ulem}
\usepackage{cancel}

\def\ba{{\bf a}}

\def\bb{{\bf b}}
\def\bc{{\bf c}}
\def\bC{{\bf C}}

\def\bd{{\bf d}}

\def\bH{{\bf H}}

\def\bQ{{\bf Q}}
\def\br{{\bf r}}

\def\bx{{\bf x}}
\def\by{{\bf y}}
\def\balpha{\boldsymbol{\alpha}}

\def\brho{\boldsymbol{\rho}}

\begin{document}

\title{Utilization of the Wavefront Sensor and Short-Exposure Images for Simultaneous Estimation of Quasi-static Aberration and Exoplanet Intensity}

\author{Richard A. Frazin\altaffilmark{1}\email{rfrazin@umich.edu}}
\altaffiltext{1}{{\it Dept. of Atmospheric, Oceanic and Space Sciences, 
University of Michigan, Ann Arbor, MI 48109}}

\begin{abstract}

Heretofore, the literature on exoplanet detection with corongraphic telescope systems has paid little attention to the information content of short exposures and methods of utilizing the measurements of adaptive optics wavefront sensors.
This paper provides a framework for the incorporation of the wavefront sensor measurements in the context of observing modes in which the science camera takes millisecond exposures.
In this formulation, the wavefront sensor measurements provide a means to jointly estimate the static speckle and the planetary signal.
The ability to estimate planetary intensities in as little as few seconds has the potential to greatly improve the efficiency of exoplanet search surveys.
For simplicity, the mathematical development assumes a simple optical system with an idealized Lyot coronagraph.
Unlike currently used methods, in which increasing the observation time beyond a certain threshold is useless, this method produces estimates whose error covariances decrease more quickly than inversely proportional to the observation time.
This is due to the fact that the estimates of the quasi-static aberrations are informed by a new random (but approximately known) wavefront every millisecond.
The method can be extended to include angular (due to diurnal field rotation) and spectral diversity.
Numerical experiments are performed with wavefront data from the AEOS Adaptive Optics System sensing at 850 nm.
These experiments assume a science camera wavelength $\lambda$\ of $1.1 \,  \mu$, that the measured wavefronts are exact, and a Gaussian approximation of shot-noise.
The effects of detector read-out noise and other issues are left to future investigations.
A number of static aberrations are introduced, including one with a spatial frequency exactly corresponding the planet location, which was at a distance of $\approx 3\lambda/D$\ from the star.
Using only 4 seconds of simulated observation time, a planetary intensity, of $\approx 1$ photon/ms, a  stellar intensity of $\approx 10^5$\ photons/ms (contrast ratio $10^5$), the short-exposure estimation method recovers the amplitudes static aberrations with a 1\% accuracy, and the planet brightness with a with 20\% accuracy.

\end{abstract}

\maketitle

\section{Introduction and Motivation}

The idea of using millisecond exposures for ultra-high contrast imaging, such as is required for exoplanet science, has been given little attention due to a variety of issues ranging from detector readout noise, to high-volume data management and undeveloped computational methodologies.   
The method described here is to record the intensity in the science camera simultaneously with the wavefront sensor (WFS) measurements of the adaptive optics (AO) residual phase and solve an inverse problem to simultaneously determine the quasi-static (QS) aberrations and the planetary emission.
Intuitively, it seems that the combination of millisecond exposures and the measurements of the WFS provide much more ability to estimate the QS aberrations and the planetary emission than do long-exposure measurements, even when combined with a scheme to estimate QS aberrations, such as a focal plane mask.
There are two main reasons for this.
The first reason is that the random AO residual phase provides a new pupil plane phase screen every millisecond, which completely de-correlates on the time-scale of the wind blowing across the telescope aperture, the so-called ``atmospheric clearing time," $\tau_c$\ (McIntosh et al.~2005).
Thus, every $\tau_c$\ seconds AO residual provides a statistically independent phase screen which modulates the quasi-static speckles in a new way.
Each new phase screen provides more diversity in the observations and gives more power to estimate the QS aberrations and the spatial distribution of planetary emissions.
In contrast, using a focal plane mask and with long-exposure images to estimate errors may introduce non-common path errors, and the focal plane mask provides only one measurement with no diversity.
However, a focal plane mask could be combined with short-exposure imaging to increase the diversity, using a methodology similar to the one advocated here.

The second reason for advocating millisecond images is that short-exposure measurements take advantage of the times when the random temporal modulation greatly reduces the intensity of the speckle background (at a given spatial point), which is the so-called so-called ``dark speckle'' principle described by Labeyrie (1995) and developed in the context of coronography by Boccaletti et al.~(1998).
The method described here can be considered a generalization of the dark-speckle method that takes advantage of both the WFS information and useful properties of the photon noise, particularly the fact that variance of the Poisson distribution is equal to its mean (i.e., there is less shot-noise when the speckle background is faint).

Roggemann and Meinhardt~(1993) used wave-front sensor measurements to deconvolve images, but did not consider the effects of quasi-static speckle, which are critical for high-contrast imaging.
Codona et al.~(2008) describe a short-exposure technique called phase sorting interferometry which utilizes the WFS data to help determine the quasi-static aberrations.
However, unlike this paper, they do not consider planetary emissions, which have some potential to be confused with quasi-static aberrations. 
Gladysz et al.~(2010) explored short-exposure statistics of stellar speckle and planetary emission, and showed that the resulting histograms provide some means to discriminate between planetary emission and speckle.   
However, reducing the intensity time-series to histograms discards much information, especially WFS' estimate of the AO residual phase, which is the function that is driving the temporal modulation in the first place.

The current state-of-the-art post-processing methods, always applied to long exposure images, attempt to subtract the point spread function (PSF), which includes the effects of QS aberrations,  by taking advantage of diversity introduced by the diurnal field rotation, multiple spectral channels and/or observations of naked stars with the telescope in a (hopefully) similar state.
The most well-known of these are called Angular Differential Imaging (ADI; Marois et al.~2006), and Locally Optimized Combination of Images (LOCI; Lafreni\`ere et al. 2007).
At the time of this writing, the latest variant along these lines utilizes principle components analysis to treat some of the difficulties of ADI and LOCI (including overly-aggressive subtraction and bias) and is called Karhunen Lo\`eve Image Projection (KLIP) algorithm (Soummer et al.~2012).
An independent line of thought in exoplanet detection involving the scanning Hotelling observer has been proposed by Caucci et al.~(2007, 2009), who did not consider the effects of QS aberrations or a coronagraph (cf. Sec.~\ref{DectEst} below).
The work here is related to that of Sauvage et al.~(2010), in which the authors proposed to use an analytical coronagraph model to simultaneously estimate the planetary emission and the stellar background from a long-exposure image.  
The most significant difference between this work and that of Sauvage et al.~(2010) is the incorporation of WFS data and short-exposure images.


\section{A Simple Coronagraph Model}\label{model}

\begin{figure}
\includegraphics[width=\linewidth]{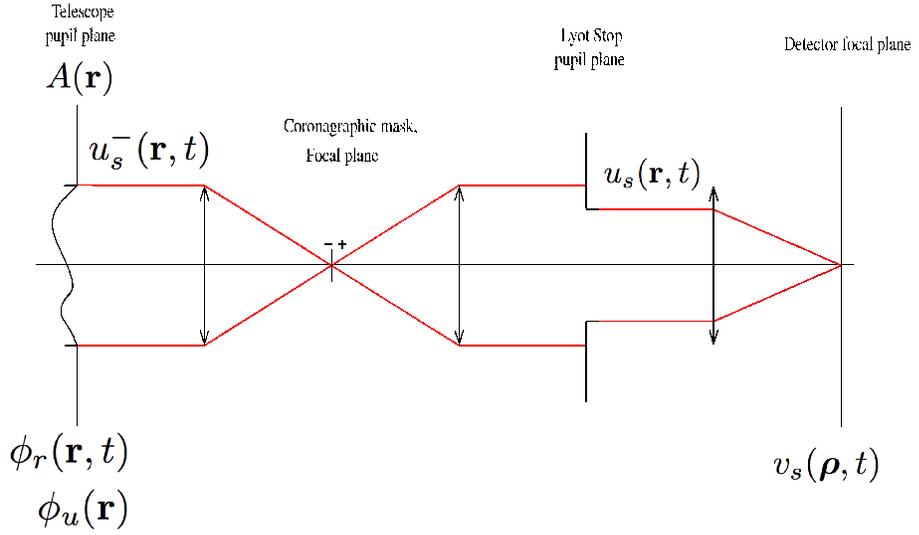}
\caption{\small (color online) Optical scheme of a coronagraphic imager.
The incoming AO residual is $\phi_r(\br,t)$, and the upstream static aberration is denoted $\phi_u(\br)$ (downstream static aberrations are not included in this analysis).
The  pupil plane fields upstream and downstream of the coronagraphic focal plane mask and downstream of the Lyot stop are given by $u_s^-(\br,t)$, and  $u_s(\br,t)$, respectively.   The aperture function $A(\br)$\ accounts for both the telescope aperture and the Lyot stop.
From Sauvage et al.~(2010).}
\label{fig_coronagraph}
\end{figure}

Following Sauvage et al.~(2010), consider a telescope with an AO system feeding a stellar coronagraph, shown schematically in Fig.~\ref{fig_coronagraph}.
Shown here is stellar variant of the classic Lyot coronagraph (Lyot~1939), versions of which have been used to study the Sun's corona for many decades.
In a Lyot coronagraph, light is first focused onto a focal-plane occulter, onto which the image of the star falls.
The light that misses the occulter is then re-collimated and passed through the so-called Lyot stop, which removes much of the star light diffracted by the focal plane occulter.

Assume an aperture function $A(\br)$, where $\br$\ is the 2D spatial coordinate in the pupil plane.  In this formulation, $A(\br)$\ is assumed to include the effect of the Lyot stop in the coronagraph system (Sauvage et al.~2010) and may have a non-zero imaginary part.
Above the atmosphere, the field due to the on-axis star is given by the (real) constant $u_*$, and the field due to the planet is given by the linear phase-angle  function $u_\bullet \exp[ j k \balpha_p \cdot \br  ]$, where $u_\bullet$\ is real and $u_\bullet/u_*$\ is the square-root of intensity contrast ratio between the star and the planet, $k = 2\pi / \lambda$\ and $\lambda$ is the wavelength at the central wavelength of observation, and $\balpha_p$\ is the 2D direction vector ($|| \balpha_p || < 1$) indicating the location of the planet.
This is easily generalized to multiple planets and extended objects (e.g., protoplanetary disks) with superposition of sources at angles $\{\balpha_k \}$.
The AO residual phase is given by $\phi_r(\br,t)$\ and the upstream (of the coronagraph) static aberrations (leading to so-called quasi-static speckle) are given by $\phi_u(\br)$.
This formulation makes the simplification of excluding time-dependence of $\phi_u$.
While this is justified on the time-scales of seconds, or possibly minutes, it is not valid for longer observation times.   
One could describe slow changes of $\phi_u$, perhaps using B-splines to parameterize the temporal variation, but this is left to future investigations.

The  total wavefront phase aberration is given by: 
\begin{equation}
\phi(\br,t) \equiv \phi_r(\br,t) + \phi_u(\br) \, .
\label{total_phase}
\end{equation}
Generally speaking one expects that $||\phi_u(\br)|| << || \phi_r(\br,t) ||$.  The importance of $\phi_u$\ is due to the fact that it changes on very long time scales compared to $\phi_r$\ and hence creates speckles that cannot be mitigated by long integrations.
For simplicity, static aberrations downstream of the coronagraph, which are generally considered to be less important than upstream aberrations, are not considered, see Sauvage et al.~(2010) for a discussion.
The analysis here permits amplitude effects by allowing $\phi_r$\ and $\phi_u$\ to be complex valued functions.
This analysis assumes that the planet and the star are in close enough proximity so that they are in the same isoplananatic patch, but are separated enough for the planet to be beyond the influence of the coronagraph, i.e., $\theta_{\mathrm{in}}<||\balpha_p||<\theta_{\mathrm{iso}}$, where $\theta_{\mathrm{in}}$\ is the inner-working-angle of the coronagraph and $\theta_{\mathrm{iso}}$\ is the isoplanatic angle.
 
Without loss of generality, assume that the spatial average AO residual phase over the pupil is zero:
\begin{equation}
\phi_{r0}(t) \equiv \int \mathrm{d}^2 \br A(\br) \phi_r(\br,t) = 0 \, ,
\label{zero_mean_residual}
\end{equation}
as the value of $\phi_{r0}$\ does not have any effect on the intensity.  Thus, the pre-coronagraph pupil plane fields of the star and planet are given by:
\begin{eqnarray}
u_s^-(\br,t) & = & u_* A(\br) \exp[ j   \phi(\br,t) ]
\label{pupil_up_star} \\
u_p(\br,t) & = & u_\bullet A(\br) \exp [ jk \balpha_p \cdot \br + j\phi(\br,t)  ] \, . 
\label{pupil_planet}
\end{eqnarray}
and the $^-$\ superscript denotes that this field is upstream of the coronagraph (see Fig.~\ref{fig_coronagraph}).  
Since the planetary light is assumed not to be effected by the coronagraph, this notation is omitted for $u_p$.

To model the effect of a the coronagraph in the optical system, one may use the formalism of Sauvage et al. (2010) , in which ``the perfect coronagraph is always defined as an optical device that subtracts a centered Airy pattern from the electromagnetic field."  The perfect coronagraph concept seems to have originated with Cavarroc et al.~(2006).     
In the pupil plane downstream of the coronagraph, the stellar field is given by
\begin{eqnarray}
u_s(\br,t)  & = &   u_s^-(\br,t) - u_* \eta(t)A(\br)  \nonumber \\
              & = &   u_* \left\{A(\br) \exp  [j  \phi(\br,t) ] - \eta(t)A(\br)  \right\}  \, , 
\label{coronagraph_field} \\
\end{eqnarray}
where $\eta(t)$\ is the complex number ($\parallel \eta(t) \parallel < 1$) that minimizes the total energy transmitted by the coronagraph, $\int  u_s(\br,t)u_s^\star(\br,t)\mathrm{d}^2 \br \,$.
It is easy to show that 
\begin{equation}
\eta(t) = 
\frac{\int \mathrm{d}^2 \br A(\br)A ^\star(\br) \exp  [- j  \phi_r(\br,t) ]}
{\int \mathrm{d}^2 \br A(\br)A ^\star(\br) } 
\label{eta}
\end{equation}
where the $^\star$\ superscript indicates complex conjugation, and the $\phi_u(\br)$\ has been ignored as it has little consequence here.
Note that as $\phi(\br,t)  \rightarrow 0$, $\eta(t) \rightarrow 1$, implying the extinction of all the stellar emission, which is the defining characteristic of a perfect coronagraph.

Under Fraunhofer diffraction, which relates image plane fields to pupil plane fields by Fourier transformation (e.g., Goodman 1968), the pupil-plane fields $u_p$ and $u_s$\ lead to the following post-coronagraph, image-plane fields for the planet and star:
\begin{eqnarray}
v_p(\brho,t)  & = &   u_\bullet
  \int \mathrm{d}^2 \br  \, A(\br)  \exp j \left[  -\frac{k}{f} ( \brho - f \balpha_p) \cdot \br  +  \phi_r(\br,t)    \right] \,  ,
\label{planet_image_field} \\
 v_s(\brho,t) & = & u_* \int \mathrm{d}^2 \br \, A(\br) 
\exp \left[ -j \frac{k}{f} \brho \cdot \br \right] \Big\{  \exp [ j\phi(\br,t)]   - \eta(t) \Big\}  
\label{star_image_field} 
\end{eqnarray}
where $\brho$ is the 2D image plane coordinate, $f$\ is the effective focal length, and $\phi_u(\br)$\ has been dropped from Eq.~(\ref{planet_image_field}),  as the planetary light scattered by the static aberrations is not important.  The instantaneous intensity in the image plane is given by the incoherent sum of the instantaneous planetary and stellar intensities:
\begin{equation}
I(\brho,t) = I_s(\brho,t) + I_p(\brho,t) \, ,
\label{total_intensity}
\end{equation}
where $I_p(\brho,t) \equiv v_p(\brho,t) v_p^\star(\brho,t)$\ and $I_s(\brho,t) = v_s(\brho,t)v_s^\star(\brho,t)$.
Using Eq.~(\ref{planet_image_field}), the planetary intensity is given by: 
\begin{eqnarray}
I_p(\brho,t)  & = & u_\bullet^2 \int \mathrm{d}^2 \br   \int \mathrm{d}^2 \br' \, 
 A(\br)A^\star(\br')      \times \nonumber \\
 && \exp j \left[  \frac{k}{f}( \brho - f \balpha_p )\cdot (\br'-\br) +  \phi_r(\br,t) -  \phi_r^\star(\br',t)   \right] \, .
 \label{planet_intensity}
 \end{eqnarray}
 Similarly, from Eq.~(\ref{star_image_field}), the star's intensity is given by:
\begin{eqnarray}
I_s(\brho,t)  & = & u_*^2 \int \mathrm{d}^2 \br   \int \mathrm{d}^2 \br' \, 
 A(\br)A^\star(\br')  \exp \left[ j \frac{k}{f} \brho \cdot (\br'-\br) \right]   \times \nonumber \\
 && \left\{ \eta(t)\eta^\star(t) + e^{j[ \phi(\br,t) -  \phi^\star(\br',t)]} - \eta(t)e^{-j\phi^\star(\br',t)} -  \eta^\star(t)e^{j\phi(\br,t)} \right\} \, ,
 \label{star_intensity}
 \end{eqnarray}
where the reader is reminded that $\phi(\br,t) = \phi_r(\br,t) + \phi_u(\br)$.
From this equation, it can be seen that the star's instantaneous intensity is a complicated function of AO residual $\phi_r(\br,t)$, the static aberrations $\phi_u(\br)$, and the complex extinction factor $\eta(t)$, which accounts for the effect of the ideal coronagraph.
In contrast, the planetary intensity is much more simple because the static aberrations were dropped and it was assumed earlier that the effect of the coronagraph is not important since $\balpha_p$\ is greater than the inner working angle $\theta_{\mathrm{in}}$.
Sauvage et al.~(2010) obtained an expression equivalent to Eq.~(\ref{star_intensity}), except their analysis also included aberrations downstream of the coronagraph.
As Sauvage et al.~(2010) were primarily concerned with long exposure images, they took the time average of Eq.~(\ref{star_intensity}), which eliminates $\phi_r(\br,t)$\ in favor of structure functions of the AO residual.  

\subsection{multiple planets and/or extended emission}\label{multiple}
 
Equation~(\ref{planet_intensity}) assumes planetary emission only from the sky-angle $\balpha_p$, but it easily generalizes to the case of extended planetary emission and multiple planets.
To account for such emission, one must sum incoherently over all of the various sources.
Let $J(\balpha)$\ be the brightness distribution on the sky (excluding the star), then the generalization of Eq.~(\ref{planet_intensity}) is:
 \begin{eqnarray}
I_p(\brho,t)  & = & \int \mathrm{d}^2 \balpha \, J(\balpha)  \int \mathrm{d}^2 \br   \int \mathrm{d}^2 \br' \, 
 A(\br)A^\star(\br')      \times \nonumber \\
 && \exp j \left[  \frac{k}{f}( \brho - f \balpha )\cdot (\br'-\br) +  \phi_r(\br,t) -  \phi_r^\star(\br',t)   \right] \, .
 \label{planet_kernel}
 \end{eqnarray}
 Henceforth, the double integral over $\br$\ and $\br'$\ in Eq.~(\ref{planet_intensity}) will be called the `` planetary intensity kernel.''

\subsection{modulation of planetary intensity and stellar speckle}

\begin{figure}
\epsfig{file=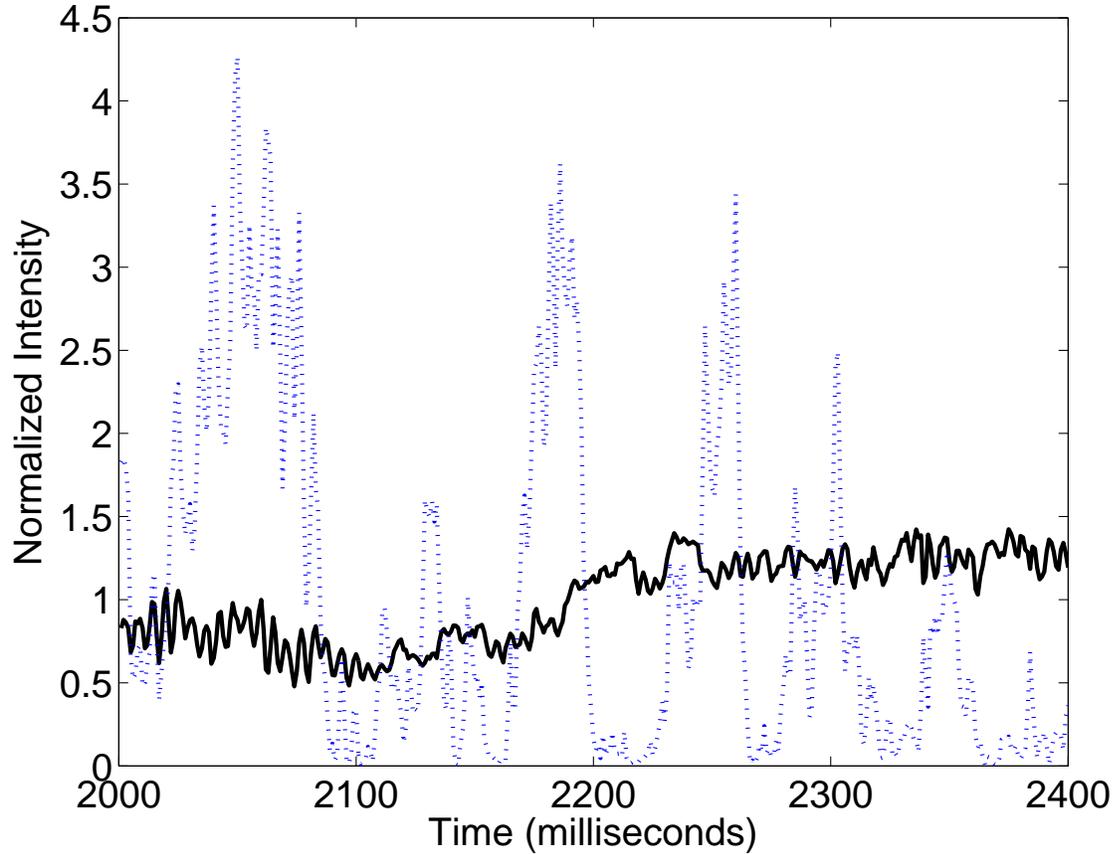,width=.9\linewidth,clip=}
\caption{\small (color online) The black solid line shows a portion of the time-series of the temporal variation of the planetary intensity from the Experiment 1 in Sec.~\protect\ref{experiments}.  The dashed blue line shows the corresponding stellar intensity at the planet's location in the image plane.  Both the planetary and stellar intensity are normalized to have a mean of unity in this figure.  The stellar intensity is enhanced by sinusoidal static aberration at the spatial frequency corresponding to planet's location (cf. Fig.~\protect\ref{fig_qsa}).}
\label{fig_modulation}
\end{figure}

\begin{figure}
\epsfig{file=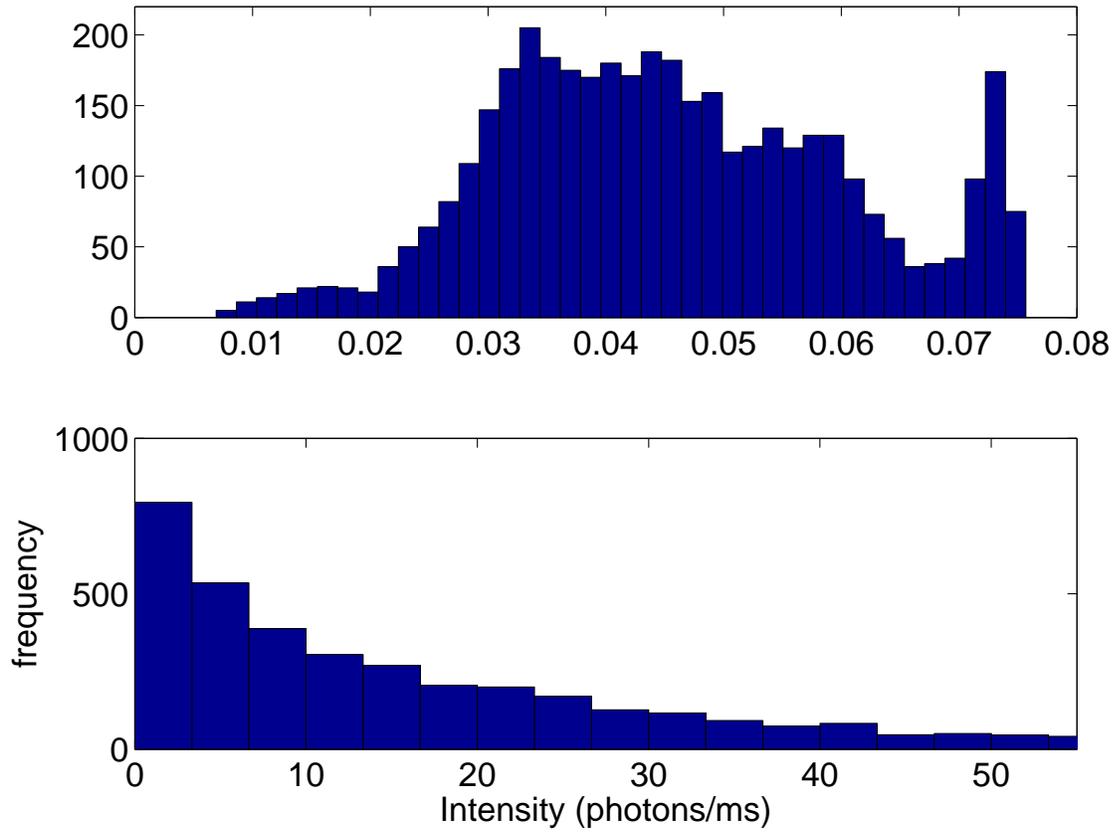,width=.9\linewidth,clip=}
\caption{\small (color online) Histograms of the planetary intensity (top) and the stellar intensity (bottom) from the full time-series
 excerpted in Fig.~\protect\ref{fig_modulation}.  These distributions exhibit the negative (top) and positive (bottom) skewness discussed in Gladysz et al.~(2010).}
\label{fig_histogram}
\end{figure}

The temporal variation of the AO residual modulates the stellar speckle and the planetary emission differently and in a way that can be exploited by short-exposure imaging.
At the planet location in the image plane, $\brho = f \balpha_p$, consider a 1\underline{st} order Taylor expansion of the exponential in Eq.~(\ref{planet_image_field}), which gives:
\begin{equation}
v_p(f \balpha_p,t)   \approx   u_\bullet \int \mathrm{d}^2 \br  \, A(\br)  [1 + j \phi_r(\br,t)]
= u_\bullet \int \mathrm{d}^2 \br  \, A(\br) \,  ,
\label{taylor_planet}
\end{equation}
where the equality makes use of Eq.~(\ref{zero_mean_residual}).
Thus, the intensity has no terms that are linear in $\phi_r$, and, to first order, the planetary field isn't modulated by the temporal variation of the AO residual phase.
Temporal variability in the planetary intensity is due to the higher order terms in the expansion.
In contrast, at the point $\brho = f \balpha_p$, Taylor  expansion  of the second exponential in Eq.~(\ref{star_image_field}) yields:
\begin{equation}
v_s(f \balpha_p,t) \approx u_* \int \mathrm{d}^2 \br \, A(\br) 
\exp   \left[  j k \balpha_p \cdot \br \right] \Big[ 1 - \eta(t) +j \phi_r(\br,t) + j \phi_u(\br) \Big]  \,  .
\label{taylor_star}
\end{equation}
One can see that Eq.~(\ref{taylor_star}) indicates that the field should be greatly modulated by the residual atmospheric fluctuations at the location of the planet.
The key differences in the temporal behavior between the planetary case [Eq.~(\ref{taylor_planet})], and the stellar  case [Eq.~(\ref{taylor_star})] is due to the spatially oscillating $\exp (- j k \balpha_p  \cdot \br) $\ factor inside the integrand, and the fact that $\eta$\ is a function of time.
Fig.~\ref{fig_modulation} gives an example, taken from Experiment 1 in Sec.~\ref{experiments}, of the difference in the modulation of the planetary vs. stellar brightness at the same point in the image plane.
In this case, the stellar brightness is enhanced by a static aberration with the spatial frequency corresponding to the planetary position.
Figure~\ref{fig_histogram} shows histograms of the planetary intensity (top) and the stellar intensity (bottom) from the same time-series.
The positive skewness of stellar distribution and the negative one of the planetary distribution was noted by Gladysz et al.~(2010), who proposed using these properties to separate planetary light from the stellar speckle.
In the case shown in Figure~\ref{fig_histogram}, the stellar intensity was a nearly perfect negative exponential distribution, corresponding to so-called ``fully developed speckle" (Goodman 2007), but other examples not shown here indicate that the stellar intensity may deviate from this distribution substantially.

The differences between the planetary and stellar modulation implied by  Eqs.~(\ref{planet_image_field}) and (\ref{star_image_field}) or Eqs.~(\ref{taylor_planet}) and (\ref{taylor_star}) hint that short exposures may provide a means to discriminate between between the planetary emission and speckle from the star.  Thus, one would expect that the WFS estimate of $\phi_r(\br,t)$ should provide some additional means to separate the planetary emission from the speckle.

\subsection{Series expansion of the QS aberrations}

Our problem requires estimation of the static aberrations, which is less difficult when they can be assumed small enough for the following linearization to be valid:
\begin{equation}
\exp [j \phi_u(\br)] \approx 1 + j \phi_u(\br) \, .
\label{linearization}
\end{equation}
This is not particularly stringent, as one expects that static errors should satisfy $|| \phi_u(\br) || << 1$\ in a good optical system.  Note that the AO residual is not assumed to be small and $e^{ j \phi_r(\br,t) }$\ is not linearized in this paper.
Calculating the instantaneous total intensity from Eq.~(\ref{total_intensity}) using Eqs.~(\ref{planet_image_field}), (\ref{star_intensity}), and (\ref{linearization})  yields:
\begin{eqnarray}
I(\brho,t)  &  = &  I_p(\brho,t)  +\mathcal{A}(\brho,t) + 
(\mathcal{B}\circ \phi_u^\star)(\brho,t) +  \nonumber \\
&& 
 (\mathcal{B}^\star \circ \phi_u )(\brho,t) +
 (\mathcal{C}\circ [\phi_u,\phi_u^\star])(\brho,t)
\label{total_intensity2}
\end{eqnarray}
where the $\circ$\ symbol indicates composition.   
Of course, this equation may be generalized to include extended objects and/or multiple planets by replacing the first term with Eq.~(\ref{planet_kernel}).
The linearization in Eq.~(\ref{linearization}) applied to Eq.~(\ref{star_intensity}), creates three types of terms:  those that do not depend on $\phi_u$, those that are linear in $\phi_u$\ and those that are quadratic in $\phi_u$.  These terms are represented, respectively, by the $\mathcal{A}$,  $\mathcal{B}$, and $\mathcal{C}$\ terms in Eq.~(\ref{total_intensity2}).
The $\mathcal{B}$\ terms lead to the so-called ``pinned speckle" in long exposure imaging (Bloemhof 2004).
Before giving formulae for these operators it is helpful to define the kernel function:
\begin{equation}
 G(\br,\br';\brho) \equiv  A(\br)A^\star(\br')  \exp \left[- j \frac{k}{f} \brho \cdot (\br-\br') \right]  \, ,
 \label{Gkernel}
 \end{equation} 
which has the useful symmetry property: $G(\br,\br';\brho) = G^\star(\br',\br;\brho)$. 

\begin{eqnarray}
\mathcal{A}(\brho,t)  &= &  u_*^2 \int \mathrm{d}^2 \br \int \mathrm{d}^2 \br' \,  G(\br,\br';\brho) \Big\{ \eta(t)\eta^\star(t) +   \exp j[ \phi_r(\br,t) - \phi^\star_r(\br',t)  ]  \nonumber \\ 
&& - \eta(t) \exp [ - j \phi^\star_r(\br',t)]  - \eta^\star(t) \exp [  j \phi_r(\br,t)]   \Big\} 
 \label{Aint}\\
  (\mathcal{B}\circ \phi_u^\star)(\brho,t) &= & u_*^2   \int \mathrm{d}^2 \br \int \mathrm{d}^2 \br' \,  G^\star(\br,\br';\brho) \phi^\star_u(\br) \times \nonumber \\
  && \Big\{  j\eta(t) \exp[-j\phi^\star_r(\br,t) ] - j\exp j[ \phi_r(\br',t) - \phi^\star_r(\br,t)  ] \Big\} 
 \label{Bint} \\
  (\mathcal{B}^\star\circ \phi_u)(\brho,t) &= & u_*^2 \int \mathrm{d}^2 \br \int \mathrm{d}^2 \br' \,  G(\br,\br';\brho) \phi_u(\br) \times \nonumber \\
  && \Big\{ - j\eta^\star(t) \exp[j\phi_r(\br,t) ] + j\exp j[ \phi_r(\br,t) - \phi^\star_r(\br',t)  ] \Big\} \\
  (\mathcal{C}\circ [\phi_u,\phi_u^\star])(\brho,t) & = & u_*^2 \int \mathrm{d}^2 \br \int \mathrm{d}^2 \br' \,  G(\br,\br';\brho) 
   \phi_u(\br)\phi_u^\star(\br') \exp j[ \phi_r(\br,t) - \phi^\star_r(\br',t)  ] \, .
\label{Cint}
\end{eqnarray}
In Eq.~(\ref{total_intensity2}), the $\mathcal{A}$\ term is due to the atmospheric residual speckles, the  $\mathcal{C}$\ term is due to the static aberrations (which are also modulated by the AO residual), and the $\mathcal{B}$\ terms are due to the mixing of these two effects.
Note that when $\phi_u =0$, $I_s(\brho,t) = \mathcal{A}(\brho,t)$.
While all of the integrals in Eqs.(\ref{Aint}) through (\ref{Cint}) can be separated to integrate over $\br$\ and $\br'$\ independently, the form given here is more conducive to presentation.

To make estimation of the QS aberrations tractable, a finite expansion for $\phi_u(\br)$ is assumed:
\begin{equation}
\phi_u(\br) = \sum_{k=1}^{K} a_k \psi_k(\br)  \,
\label{phi_expansion}
\end{equation}
where $K$\ is the number of terms required and $\{  \psi_k(\br) \}$\ are basis functions on which the static aberrations can be represented.
Examples of possible expansion bases include Fourier, Zernicke, and B-splines (e.g., Unser 1999), which include pixel bases.   
Importantly, orthogonality of the  $\{  \psi_k(\br) \}$\  will not be assumed, allowing considerable freedom in the form of the expansion.
For convenience, the quantities $\{ a_k \}$\ are placed into a $1 \times K$\ (column) vector $\ba$. 
Substituting Eq.~(\ref{phi_expansion}) into Eq.~(\ref{total_intensity2}) gives three types of terms, those that are independent of $\ba$, linear in $\ba$, and quadratic in $\ba$:
\begin{eqnarray}
I(\brho,t)  &  = & u_\bullet^2 i_p(\brho,t)  + \mathcal{A}(\brho,t) + \nonumber \\
&&  \ba^\mathrm{H} \bb(\brho,t)  + \bb^\mathrm{H}(\brho,t) \ba  + \ba^\mathrm{H}  \bC(\brho,t) \ba \, ,
\label{total_intensity3}
\end{eqnarray}
where the $^\mathrm{H}$\ superscript indicates the transpose and complex conjugate (Hermitian conjugate), $i_p(\brho,t) \equiv I_p(\brho,t)/u_\bullet^2$\ is the planetary PSF.
Eq.~(\ref{total_intensity3}) expresses the instantaneous intensity in terms of the unknown planetary intensity $u_\bullet^2$\ and the static aberration coefficients $\ba$.   
The column vector $\bb(\brho,t)$\ has $K$\ components, and $\bC(\brho,t)$\ is a Hermitian symmetric, $K \times K$\ matrix.
Thus, for fixed $(\brho,t)$, $\ba^\mathrm{H} \bb(\brho,t)$\ is the scalar product of the two complex vectors.
Similarly, $\bC(\brho,t) \ba$\ is the standard matrix-vector multiplication, with $\ba^\mathrm{H}  \bC(\brho,t) \ba$\ resulting in a scalar.
Using Eqs.~(\ref{Bint}), (\ref{Cint}) and (\ref{phi_expansion}), the components of $\bb$\ and $\bC$\ are given by:

\begin{eqnarray}
  [\bb(\brho,t)]_k &  = &  u_*^2 \int \mathrm{d}^2 \br \int \mathrm{d}^2 \br' \,  G^\star(\br,\br';\brho)  \psi^\star_k(\br) \times \nonumber \\
  && \Big\{  j\eta(t) \exp[-j\phi^\star_r(\br,t) ] - j\exp j[ \phi_r(\br',t) - \phi^\star_r(\br,t)  ] \Big\}
  \label{bvec} \\
\left[ \bC(\brho,t) \right]_{kl}  & = & u_*^2 \int \mathrm{d}^2 \br \int \mathrm{d}^2 \br' \,  G(\br,\br';\brho) \psi_k(\br)\psi_l^\star(\br') \times \nonumber \\
&&  \exp \Big\{ j \phi_r(\br,t) - j\phi^\star_r(\br',t)  \Big\}  \, .
  \label{Cmat} 
 \end{eqnarray}
 Note that Eqs.~(\ref{bvec}) and (\ref{Cmat}) are separable and can be computed rapidly with fast Fourier transforms (FFTs), as can $\mathcal{A}(\brho,t)$.

\section{Detection/Estimation Problem}\label{DectEst}

The problem of detecting and estimating the planetary emission in the framework presented here can be posed as separating the planetary and stellar emission given noisy measurements of $I(\brho,t) $\ and the WFS data, summarized as the estimate of the AO residual $\hat{\phi_r}(\br,t)$.  It follows that using the WFS to estimate $\phi_r(\br,t)$\ is a key aspect of the proposed methodology.\footnote{Alternatively, one may also consider ignoring the WFS' estimate of $\hat{\phi_r}(\br,t)$, and instead, find suitable estimators for $\mathcal{A}$, $\mathcal{B}$, and $\mathcal{C}$ directly from the WFS data.}
While some current WFS' estimate other quantities, such as $\nabla \phi_r(\br,t)$ in the case of Shack-Hartmann sensors, estimating $\phi_r(\br,t)$\ from the WFS data is always possible with additional processing.
Optimal estimation of $\phi_r(\brho,t)$ with the maximum likelihood method is discussed in (Barrett et al.~2007) and (B\'echet et al.~2009).

The most general case allows the planetary emission to be a function of the sky-angle $\balpha$, and then the problem would be estimate the spatial function $J(\balpha)$ simultaneously with the QS aberration coefficients $\ba$, a problem which falls under the category of estimation theory.
In contrast, the problem of trying to decide whether or not there is planetary emission falls under the regime of detection theory.
Considering the problem of exoplanet detection within the framework of classical detection theory, Caucci et al. (2007) simulated the scanning Hotelling observer (SHO).
The SHO is an example of a ``scanning detector," in which one considers the probability of detection of a source at one location $\balpha_k$\ at a time.   
The process of evaluating the a series of locations is called ``scanning."
In Caucci et al. (2009), they generalized their previous analysis to include a time-series
 of images.   
The possibility of multiple planets considerably complicates scanning observer framework and it is not suited for estimation of extended emission.
The, perhaps na\"ive, first step approach in this paper is to assume that emissions other located outside of the scanning template, i.e., the group of pixels simultaneously searched for planetary emission, do not significantly influence the estimate of the planetary emission or its error bar.
The numerical examples in Sec.~\ref{experiments} support this notion.
With this approach, planetary emission that is estimated to be non-zero with sufficient statistical confidence can be considered a detection.

\subsection{Linear System Formulation}\label{linear}

Here, Eq.~(\ref{total_intensity3}) is put into the canonical form for linear systems, $\by = \bH \bx$, where  $\by$ represents the ``observation" vector, $\bx$\ is the vector of unknowns and $\bH$\ is the system model or ``data," as it is sometimes called in the statistics literature.

While Eq.~(\ref{total_intensity3}) has only $K+1$, unknowns, it is nonlinear due to the quadratic term, $\ba^\mathrm{H}  \bC(\brho,t) \ba $, which may introduce numerical difficulties.
In exchange for creating an additional set of unknowns, which will increase the uncertainty in the resulting estimates, one may remove this nonlinearity.
The new $K^2$\ unknowns are the quanties $\{a_k a^\star_l\}$, which are placed into the $1 \times K^2$ vector $\bd$.  Similarly rearranging the matrix elements $\bC(\brho,t)$\ into the $ K^2 \times 1$\ vector $\bc^\mathrm{H}(\brho,t)$, one obtains:
\begin{equation}
\ba^\mathrm{H}  \bC(\brho,t) \ba = \bc^\mathrm{H}(\brho,t) \bd \, .
\label{Cc}
\end{equation}

Consider a spatial template with $N$\ locations $\{\brho_1, ..., \brho_{N}\}$, and take $\brho_1 = f \balpha_p$\ to be the location at which one desires to know whether or not a planet is present and how bright it might be.
It is assumed that the science camera exposures are simultaneous with the output of the AO system wavefront measurements, with $T$\ exposures taken at times $\{t_1, ..., t_{T}\}$.
These exposure times form the temporal template.
Thus, there are $NT$\ observations to be placed into the vector:
\begin{equation}
\by = \left[ \begin{array}{c}
\by_1 \\
\vdots \\
\by_T 
\end{array} \right] \, ,
\label{y}
\end{equation}
where the $\{\by_i\}$\ vectors (one for each exposure) are given by:
\begin{equation}
\by_i = 
\left[ \begin{array}{c}
I(\brho_1,t_i) \\
\vdots \\
I(\brho_N,t_i) \\
\end{array} \right]
-
\left[ \begin{array}{c}
\mathcal{A}(\brho_1,t_i) \\
\vdots \\
\mathcal{A}(\brho_N,t_i) \\
\end{array} \right] \, .
\label{y_i}
\end{equation}

\noindent Similarly, the (column) vector $\bx = [u_\bullet^2, \ba^\mathrm{T}, \ba^\mathrm{H},\bd^\mathrm{T}]^\mathrm{T}$, where $^\mathrm{T}$\ indicates transposition (without complex conjugation).
Following the same notation scheme, 
\begin{equation}
 \bH = 
 \left[ \begin{array}{c}
 \bH_1 \\
 \vdots \\
 \bH_T 
 \end{array} \right] \, ,
\label{H}
\end{equation} 
where the individual matrices $\{\bH_i\}$\ have the structure:
 \begin{equation}
 \bH_i = 
 \left[ \begin{array}{c c c c}
i_p(\brho_1,t_i)& \bb^\mathrm{T}(\brho_1,t_i)&\bb^\mathrm{H}(\brho_1,t_i)&\bc^\mathrm{H}(\brho_1,t_i) \\
\vdots               &   \vdots                    & \vdots                     & \vdots                \\
i_p(\brho_N,t_i)& \bb^\mathrm{T}(\brho_N,t_i)&\bb^\mathrm{H}(\brho_N,t_i)&\bc^\mathrm{H}(\brho_1,t_i) \\
\end{array} \right] 
\label{H_i}
\end{equation}

\noindent Then, under the linear formulation, the estimation problem is to find the values of $u_\bullet^2$, $\ba$\ and $\bd$ which best satisfy: 
\begin{equation}
\by = \bH
\left[ \begin{array}{l }
u_\bullet^2 \\
\ba \\
\ba^\star \\
\bd 
\end{array} \right]
\label{linear_system}
\end{equation}
\noindent In the classical form for linear systems of equations, Eq.~(\ref{linear_system}) expresses the relationship between measured instantaneous intensities, which are one of the terms in $\by$, and the unknown planetary intensity $u_\bullet^2$, the static aberration coefficients $\ba$, and the products thereof, contained in $\bd$.
Note that the scale of the linear computation may not be especially daunting, as there are only $K^2 + 2K + 1$\ unknowns, which are much less than the $NT$\ data points.  For example, the matrix $\bH^\mathrm{H}\bH$\ only has $K^2 + 2K + 1$\ singular values.

When considering spatially extended planetary emissions, the $\bx$ vector is expanded so that $u_\bullet^2$\ is replaced a column vector representing the extended emissions and  $i_p(\brho_1,t)$\ is replaced by a matching row vector accounting for the convolution with the planetary intensity kernel [cf. Eq.~(\ref{planet_kernel})].

If $\ba$\ is complex, then the estimate can be improved by demanding consistency between the estimates of $\ba$\ and $\ba^\star$, or, re-expressing the system in terms of real and imaginary parts.   Of course, if $\ba$\ is assumed to be real, meaning static errors only affect the phase, not the amplitude,  $\bx$\ simplifies to $\bx  = [u_\bullet^2, \ba^\mathrm{T},\bd^\mathrm{T}]^\mathrm{T}$, and the 2\underline{nd} and 3\underline{rd} columns of $\bH$\ are summed together to form a single column.
In addition, a positivity constraint may be applied to the planetary intensity.

\subsection{Quadratic System Formulation}\label{quadratic}

As explained at the beginning of Sec.~\ref{linear}, the linear system in Eq.~(\ref{linear_system}) requires $K^2+K$\ aberration coefficients instead of only $K$, and both formulations have the same input data from which these parameters must be estimated. 
It follows that the quadratic formulation will produce more accurate estimates with tighter error bars, simply because the number of quantities that must jointly estimated is much smaller.
The only reason not to employ the quadratic formulation is due to the numerical difficulties that may be encountered when solving the system and estimating the uncertainties. 
The linear formulation may improve the computational possibilities for the quadratic system by providing a ``warm start" (i.e., a good initial guess).
It may be useful to note that given linear estimates of $\ba$\ and $\bd$, $\hat{\ba}$\ and $\hat{\bd}$, and their error covariances $C_{\ba \ba}$, $C_{\ba \bd}$\ and $C_{\bd \bd}$, in a second computation, one could improve the estimates of $\ba$\ by using the knowledge that each component of $\bd$\ is equal to $a_i a_j^\star$\ for a particular choice of $i$\ and $j$.
These improved estimates may provide a better warm start for the quadratic system solver. 

The objective of remainder of this section is to write the system in the form $\by = \bH'\bx' + \bQ(\bx')$, where $\bQ$\ is the quadratic term.  The vector $\by$\ is the same in both the linear and quadratic formulations. 
The quadratic version of $\bH$, $\bH'$\ is the same as $\bH$, except with the final column removed:

\begin{equation}
 \bH_i' = 
 \left[ \begin{array}{c c c }
i_p(\brho_1,t_i)& \bb^\mathrm{T}(\brho_1,t_i)&\bb^\mathrm{H}(\brho_1,t_i) \\
\vdots               &   \vdots                    & \vdots                       \\
i_p(\brho_N,t_i)& \bb^\mathrm{T}(\brho_N,t_i)&\bb^\mathrm{H}(\brho_N,t_i)  \\
\end{array} \right] \, ,
\label{HQ}
\end{equation}
so that $\bH' = [\bH_1^\mathrm{T}, ... , \bH_T^\mathrm{T}]^\mathrm{T}$.
\noindent Similarly, $\bx' = [u_\bullet^2, \ba^\mathrm{T}, \ba^\mathrm{H}]^\mathrm{T}$.

To put the quadratic term in a form suitable for digital computation, first note that for each of the $NT$\ points in the spatio-temporal template one must calculate the scalar value of $ \ba^\mathrm{H}\bC(\brho,t)\ba$, which is achieved with the following formulation:
\begin{equation}
\bQ(\ba) \equiv
\left(  \boldsymbol{1}_{NT}  \otimes \ba^\mathrm{H} \right)
\left[ \begin{array}{l}
\bC(\brho_1,t_1) \\
\vdots \\
\bC(\brho_N,t_1) \\
\bC(\brho_1,t_2) \\
\vdots \\
\bC(\brho_N,t_T)
\end{array}\right]
\ba  
\label{QQ}
\end{equation}
where $\boldsymbol{1}_{NT}$\ is the row vector $[1,...,1]$\ with $NT$\ components, and $\otimes$\ is the Kronecker matrix product.
This Kronecker product creates a large row vector with $NT$\ copies of the row vector $\ba^\mathrm{H}$.
Then, the quadratic variant of the problem can be stated as finding the $u_\bullet^2$\ and $\ba$\ which best satisfy:
\begin{equation}
\by = \bH' 
\left[ \begin{array}{l }
u_\bullet^2 \\
\ba \\
\ba^\star
\end{array} \right] + 
 \left(  \boldsymbol{1}_{NT}  \otimes \ba^\mathrm{H} \right)
\left[ \begin{array}{l}
\bC(\brho_1,t_1) \\
\vdots \\
\bC(\brho_N,t_1) \\
\bC(\brho_1,t_2) \\
\vdots \\
\bC(\brho_N,t_T)
\end{array}\right]
\ba    \, .
 \label{quad_system}
 \end{equation}

\noindent Eq.(\ref{quad_system}) is the quadratic analog of Eq.(\ref{linear_system}).

\subsection{Uncertainty Propagation}\label{UncertProp}

This section considers uncertainty propagation within the linear formulation of the problem, $\by = \bH \bx$ provided in Sec.~\ref{linear}.
The quadratic version of the problem, formulated in Sec.~\ref{quadratic}, will have smaller uncertainties due to the much smaller number of variables to estimate, but determining them requires more specialized mathematical procedures than the linear case.

This section will assume that $NT$\ is much greater than the number of parameters to be estimated.
In practice this will be satisfied, as millisecond exposure times will produce a lot of data.
For example, with a spatial template consisting of 1000 pixels, and exposure times of $10^{-3}$\ seconds, one hour of observation gives $NT = 3.6\times10^9$\ data points.
If the QS aberration expansion has, say, 20 terms, the $\bx$ vector will have 441 components, and the linear problem can be considered with the framework of least-squares.

The first term in $\by$, the observed intensity $I$,  inherits its statistical properties from noise in the observations (most likely dominated by Poisson counting noise and detector read noise), which are treated with relative ease.
However, it is of critical importance to bear in mind that the errors in $\hat{\phi_r}(\br,t)$\  will cause uncertainties in both $\by$\ and $\bH$\ which must be propagated to $\bx$ [cf. Eqs.~(\ref{y}) and (\ref{H})].
Standard least squares is likely to be insufficient, and total least squares (TLS) is a more appropriate analysis framework (Van Huffel and Vanderwalle 1991). 

The short-exposure approach proposed here is unique among the various processing methods in that knowledge of the quasi-static aberrations increases with the observation time, whereas other methods are limited by the systematic error of imperfect {\it a-priori} knowledge of the PSF.
As the condition of the matrix $\bH$\ improves with each random wavefront, increasing observation time reduces the covariance of $\hat{\bx}$ more rapidly than $1/T$.
This can be shown for the simple case of a perfect WFS, for which $\hat{\phi_r}(\br,t) = \phi_r(\br,t)$.   This implies that $\bH$\ is known exactly, as is $\mathcal{A}(\brho,t)$, so that the only uncertainty remaining comes from the measurement noise in $I(\brho,t)$, which is manifested in $\by$\ according to Eq.~(\ref{y_i}).   Consider a spatial template of $N$\ detector pixels and assume that the measurement noise of each exposure has an $N\times N$\ diagonal covariance matrix $\Sigma_N$, which is the same for each of the $T$\ exposures.
 Then, the total covariance matrix for all of the $NT$\ data points is given by large diagonal matrix $\Sigma = \mathrm{I}_T  \otimes \Sigma_N $, where $\mathrm{I}_T$\ is the $T$-by-$T$ identity matrix, and log-likelihood function is therefore given by:
 \begin{equation}
 \ln \mathrm{L}( \bx | \by) \propto - (\by - \bH \bx)^\mathrm{H} \Sigma^{-1} (\by - \bH \bx) \, .
 \label{log-likelihood}
 \end{equation} 
 This likelihood is maximized by $\hat{\bx} = (\bH^\mathrm{H}\Sigma^{-1}\bH)^{-1}\bH^\mathrm{H}\Sigma^{-1}\by$, and the covariance of $\hat{\bx}$ is given by the $(K^2 + 2K +1)$-by-$(K^2 + 2K+ 1)$\ matrix:
 \begin{equation}
 \Sigma_\bx = \left( \bH^\mathrm{H}\Sigma^{-1} \bH \right)^{-1} =
   \left( \sum_{i=1}^{T} \bH_i^\mathrm{H}\Sigma_N^{-1} \bH_i   \right)^{-1} \, .
 \label{Sigma}
 \end{equation}
Now, each of the matrices $(\bH_i^\mathrm{H}\Sigma_N^{-1} \bH_i)$\ is positive semi-definite, and the trace (which is the sum eigenvalues) of the sum $\sum_{i=1}^{T} (\bH_i^\mathrm{H}\Sigma_N^{-1} \bH_i)  $\ must increase, on average, linearly with the number of exposures $T$.
Furthermore, since each of the $\bH_i$\ is a manifestation of an independent, random phase screen $\phi_r(\br,t_i)$, the matrix
$[\bH_1^\mathrm{T},...,\bH_{i+1}^\mathrm{T}]^\mathrm{T}$, will have an equal or, more likely, smaller ratio of largest-to-smallest singular values than the matrix $[\bH_1^\mathrm{T},...,\bH_{i}^\mathrm{T}]^\mathrm{T}$\ (in other words, the condition of $\bH$\ improves with increasing observation time).   
Thus, one has: 
\begin{equation}
 \Sigma_\bx =   \left(   \sum_{i=1}^{T}  \bH_i^\mathrm{H}\Sigma_N^{-1} \bH_i  \right)^{-1}  < \left( T \left<   \bH_i^\mathrm{H}\Sigma_N^{-1} \bH_i   \right> \right)^{-1} \, ,
 \label{average}
\end{equation} 
where the brackets $< \, >$\ indicates ensemble average over the statistics of the AO residual.
Thus, it is clear that $\Sigma_\bx$\ decreases more quickly than $1/T$.  This better than $1/T$ performance was observed in the numerical experiments in Section \ref{experiments}, which compared simulated observations of 1 and 4 s.
Demonstrating the inequality Equation~(\ref{average}) in a rigorous manner is beyond the scope of this paper, but the mathematically inclined reader may consult Tropp (2012).    
While generalization of this proof to the case in which $\bH$\ also has uncertainty and its own covariance matrix is likely to be substantially more difficult, it seems that this basic result is unlikely to change.

\section{Numerical Simulations}\label{experiments}

This section explores some of the concepts described above with numerical simulation.
In the experiments described below, the residual phase data taken by the WFS on the AEOS Adaptive Optics System (Roberts \& Neyman 2002), interpolated onto an $80 \times 80$\ grid with a circular aperture inscribed (the radius $D/2$ of the circle corresponding to the telescope aperture was 40 pixels), was used as $\phi_r(\br,t)$.
The interpolation was done with a standard spline interpolation routine.  
In this way, realistic wavefronts were used.
By taking the WFS data as the true wavefront, the problem imperfect knowledge of the wavefronts was eliminated in these proof-of-principle experiments.
Including the effects of imperfect wavefront knowledge is the subject of a forthcoming study.   
AEOS has 941 actuators and the wavefront data utilized here corresponds to a sampling rate of $10^{-3}$\ s.
The AEOS sensing wavelength was 850 nm and these numerical exercises assumed a science camera operating at $1.1 \,  \mu$\ that samples at exactly the rate as the WFS (with no delay).
The residual phase value was converted from the 850 nm value to the $1.1 \,  \mu$\ value by assuming that the optical path length difference is independent of wavelength, so that the residual phase in radians simply scales inversely with the wavelength.  
In reality, the 1.1 $\mu$\ wavefront would have amplitude errors and non-common path errors, but such effects were not considered.

\begin{figure}
\epsfig{file=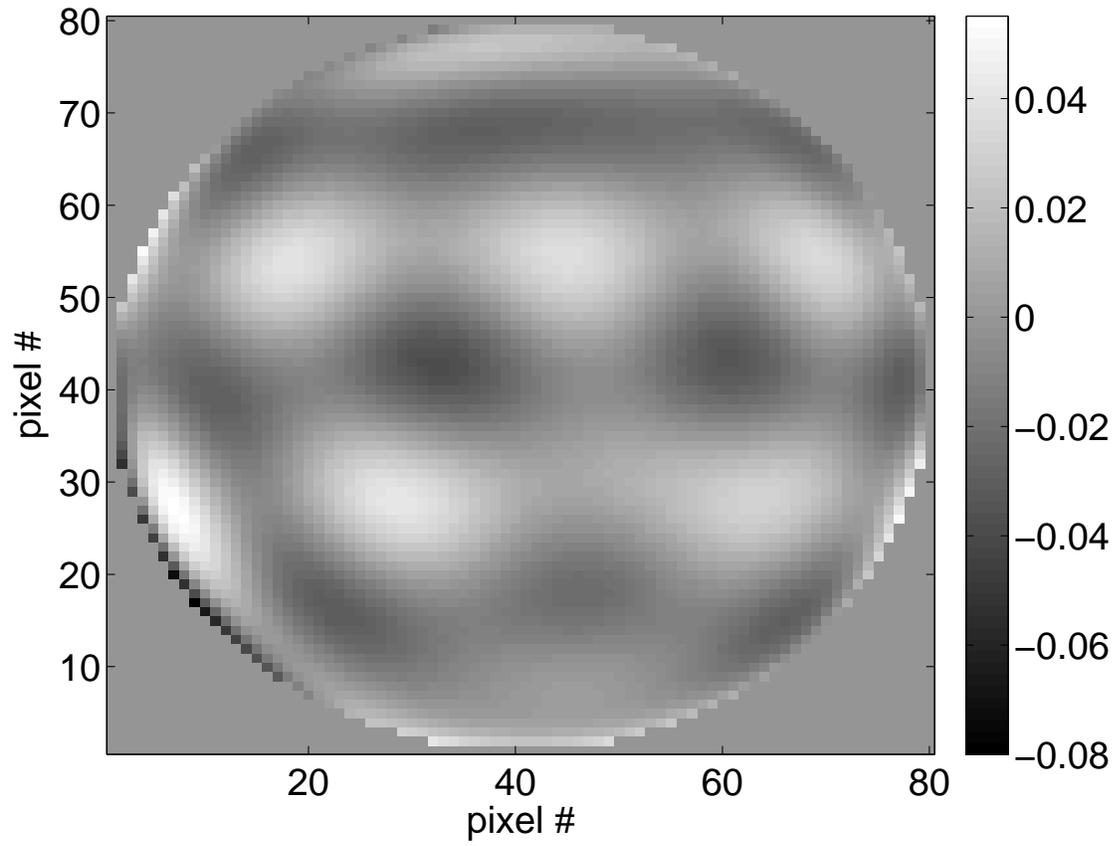,width=.9\linewidth,clip=}
\caption{\small The static aberration function assumed for all experiments.  This function is composed of a linear combination of high-order Zernike polynomials plus a sinusoid corresponding to the spatial frequency of the planet's location.}
\label{fig_qsa}
\end{figure}

\begin{figure}
\epsfig{file=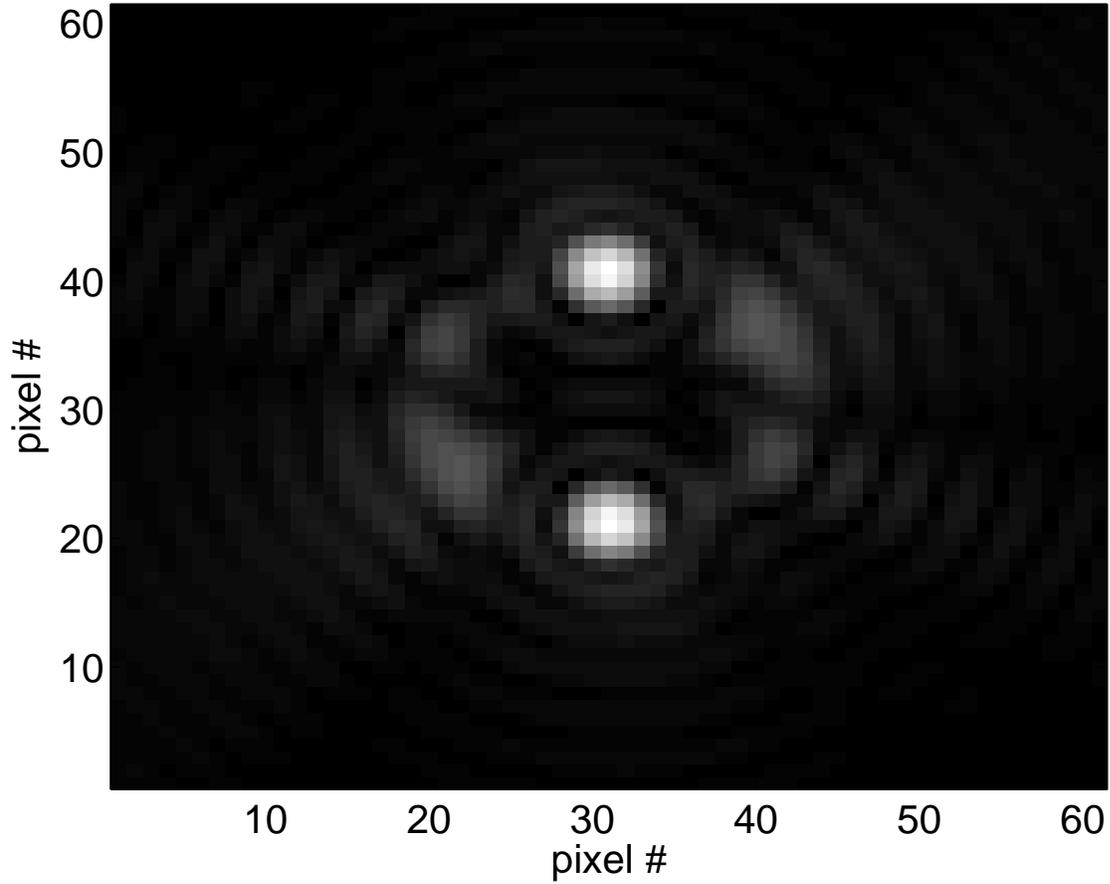,width=.9\linewidth,clip=}
\caption{\small The Fourier transform magnitude of static aberration function shown in Figure~\ref{fig_qsa}.  The two bright spots, the lower of which is spatially coincident with the planet (at distance from the star of 3$\lambda/D$), are due to the sinusoidal aberration.  The zero-padding and other details are described in the text.  The pixel numbers correspond to the central $61\times 61$\ pixels of the zero-padded array.}
\label{fig_ftqsa}
\end{figure}

As per Equations~(\ref{pupil_up_star})~and~(\ref{pupil_planet}), the planetary and stellar pupil plane fields were subjected to the same static aberrations as well as the random AO residual.   The static aberration was also placed onto the $80 \times 80$\ grid with a circular aperture inscribed.
The static aberrations included 11 terms in the expansion shown in Eq.~(\ref{phi_expansion}):  Zernike functions numbered 41-50 as well as a sinusoid which places a speckle at the planetary location as well as its conjugate point.
The coefficients $\{a_k\}$ as well as the aberration functions $\{\psi_k\}$ were real, corresponding to phase-only aberrations. 
The chosen values of $\{a_k\}$ resulted in the aberration function shown in Figure \ref{fig_qsa}.
All Fourier transforms were calculated by zero-padding the $80 \times 80$\ arrays into a $256 \times 256$\ arrays and then using the FFT.  Figure \ref{fig_ftqsa}\ shows the central portion of the so-calculated magnitude of the Fourier transform of static aberrations shown in Figure~\ref{fig_qsa}. 
The two bright points are the result of the sinusoidal aberration, and the bottom one is exactly spatially coincident with the planet location in Experiments 1 and 2 below. 
With the aperture and zero-padding as described above, the minimum of the Airy function ($ \approx 1.22 \lambda/D$\ when calculated analytically) is about 4 pixels, so, in these simulations, 1 pixel (in the image plane) corresponds to an angle of about 0.3 $\lambda/D$.
The bottom spot in Figure~\ref{fig_qsa} and the planet are 10 pixels away from the center of the image (the star's location), corresponding to a distance of $\approx 3 \lambda/D$.

The results of 3 experiments, numbered 1, 2 and 3, are given below.  The purpose of Experiment 1 is to demonstrate the ability to detect a planet under the above conditions with an assumed position (pixel location).
Treating the more common case of an unknown planet position would require a combination of generalizing the formulation to consider emission in a group pixels, called the ``scanning template", as described near the end of Sec.~\ref{linear}, and, in a series of subsequent calculations, moving the scanning template to cover the required region, as described near the beginning of Sec.~\ref{DectEst}.
The experiments here correspond to a scanning template consisting of a single pixel.
Experiment 2 identical is to experiment 1, except that it uses only one-quarter of the exposures in Experiment 1.
The purpose is, upon comparison to Experiment 1, to provide a demonstration of the claim in Equation~\ref{average}, which is that the estimate variance decreases faster than than $1/T$, where $T$\ is the total observation time.
Experiment 3 concerns the viability of the scanning concept.   
Ideally, emission from outside the scanning template would not influence the estimation of emission from within the template, but, in practice, there will be leakage.
This experiment gives an indication that leakage is unlikely to be a severe problem.
All experiments used the linear formulation in Section~\ref{linear}.
Presumably, utilizing the quadratic formulation of the problem described in Sec.~\ref{quadratic} would lead to superior results (due to the smaller number of unknowns) than those reported here, but the difficulties in minimizing the quadratic form in Eq.~(\ref{quad_system}) and in determining the uncertainties in the estimates are left to a later study.

\begin{figure}
\epsfig{file=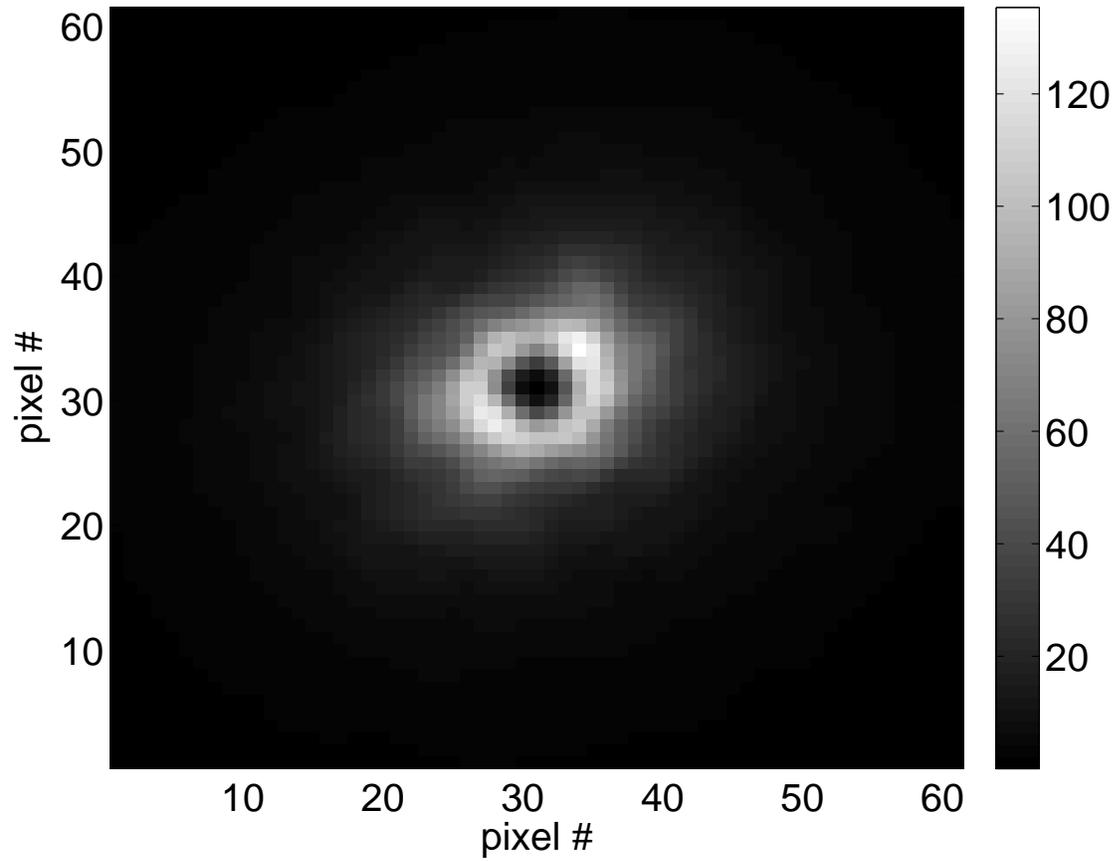,width=.9\linewidth,clip=}
\caption{\small The stellar intensity averaged over all 4000 exposures in Experiment 1 without noise.  The planet is not included and it is too faint to be seen.  The colorscale is in units [photons/ms/pixel].  The pixel numbers correspond to the central $61\times 61$\ pixels of the zero-padded array.}
\label{fig_avstar}
\end{figure}

These experiments utilized either 1 s (1000 exposures) or 4 s (4000 exposures) of AEOS data.
Figure \ref{fig_strehl} shows the $1.1 \,  \mu$\ Strehl ratio for all 4000 exposures, given by $| \eta(t) |$\ in Eq.~(\ref{eta}).
Also depicted on the same graph is the fraction of the stellar emission transmitted by the coronagraph.
In the very best frames, it removes about 95\% of the stellar emission and much less when the Strehl ratio falls.
The 850 nm Strehl ratios are smaller and are not shown here.   
The contrast ratio between the planetary and stellar brightness was $10^{-5}$, and the stellar intensity about $10^5$ photons/ms.
Integrated over the entire image plane, the time-average planetary intensity was about $1$\ photon/ms, however, at the pixel corresponding to the center of the planet's Airy pattern, the intensity was only about $0.045$\ photons/ms/pixel.
The average stellar intensity at that pixel (enhanced by the sinusoidal aberration) was  over 500 times greater, with an intensity of over $23$\ photons/ms/pixel.
Integrated over the image plane, the time-average stellar intensity was about $4.5 \times 10^4$\ photons/ms, meaning that, on the average, the coronagraph killed about $55\%$\ of the star light.
The noise-free stellar brightness, averaged over all 4000 frames, is shown Figure~\ref{fig_avstar}, where planet is too faint to be seen.

The numerical experiments here simulate the shot-noise that is associated with photon detection, but do not include detector noise, which is left to a forthcoming study.
For the purposes of the signal modeling in this  paper, the semi-classical model for photoelectric detection (Goodman 1985) is adequate.  
In the semi-classical photo-detection paradigm, there is a classical intensity (i.e., not quantized) incident upon the detector, but the detector measurement is modeled as a Poisson distributed random variable with the mean equal to the classical intensity.\footnote{The units of the classical intensity must be the $[\mathrm{photons}/\mathrm{time}] \times [\mathrm{exposure \, time}]$. }
In order to avoid the complications associated with estimation in non-Gaussian noise, Gaussian noise was chosen to behave similarly to the Poisson distribution.  Thus, if the (noise-free) intensity was $I$\ photons in a millisecond exposure, the ``measured" value (i.e., the simulated measurement) would be $I + n$, where $n$\ was taken from a zero-mean normal distribution with a variance equal to $I$. 
This Gaussian shot-noise was added to each pixel in each frame, constituting a single noise realization.
Additional noise realizations within the same experiment used same intensity values $I$, but different random values of $n$.  
The experiments were run multiple times, with different noise realizations in order to compare the variances in the estimates with the analytically derived variances from Equation~(\ref{Sigma}).

As per Equation~(\ref{y_i}), a spatial template $\{\brho_1,...,\brho_N\}$ was chosen.  For these experiments, this template corresponded to the central $27\times 27$\ pixels of the $256\times 256$\ zero-padded array in which the Fourier transforms were calculated.  This template choice excluded 30\% of the stellar photons and 20\% of the planetary photons from analysis (this difference is probably due to the coronagraph), resulting in an $\bH$\ matrix (for the 4000 exposure case) with $4000\times27^2   = 2916000$\ rows and $1 + 11 + 11^2=133$\ columns.

\begin{figure}
\epsfig{file=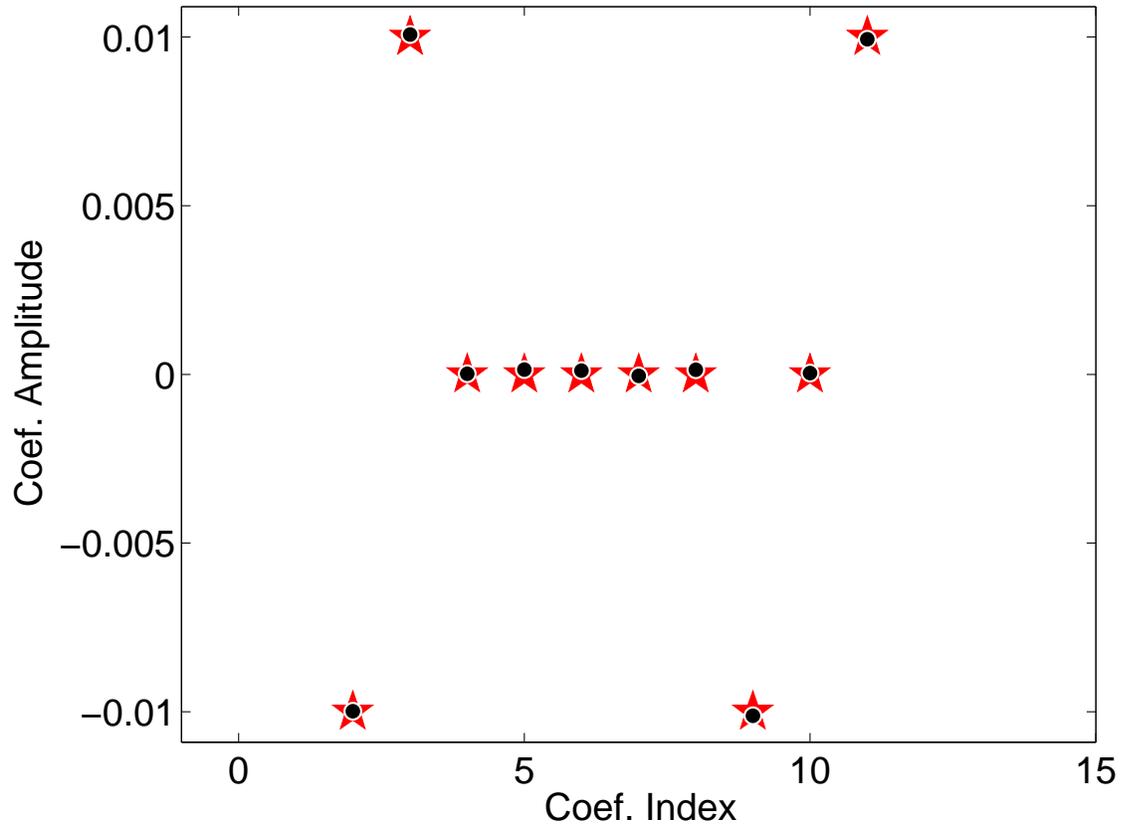,width=.9\linewidth,clip=}
\caption{\small (color online) Static aberrations coefficients $\{a_k\}$.  The red pentagrams represent the true values used in the simulation, resulting in Fig.~\protect\ref{fig_qsa}.  The black circles represent the values estimated by the algorithm in Experiment 1.  The error bars, with values of about $1\times10^{-4}$, are smaller than the plotting symbols.}
\label{fig_lincoef}
\end{figure}

\begin{figure}
\epsfig{file=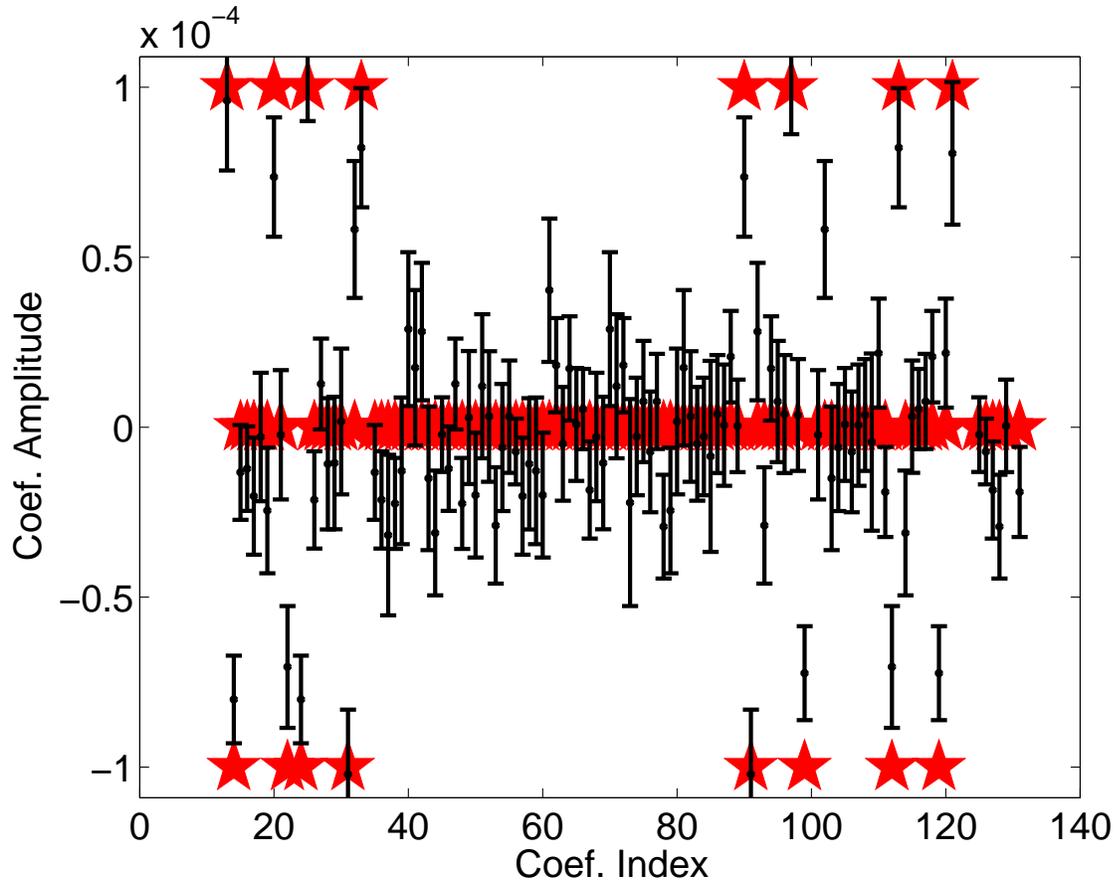,width=.9\linewidth,clip=}
\caption{\small (color online) Products of aberrations coefficients $\{a_j a_k\}$.   These quantities need not be estimated in quadratic formulation (Section~\ref{quadratic}). The red pentagrams represent the true values used in the simulation.  The black circles represent the values estimated by the algorithm in Experiment 1.  The error bars are the square-roots of the diagonal elements of the covariance of the linear estimator.}
\label{fig_quadcoef}
\end{figure}

Experiment 1 included 4000 exposures, with a mean Strehl ratio of 0.75, and was run 10 times.  
The only difference between the runs was the realization of the noise, with each noise realization consisting of 2916000 random numbers.
These 10 runs had a mean estimate of the planetary intensity of 1.1 photons/ms and a standard deviation of 0.16, which is comparable to the analytical result [i.e., the square-root of the 1-1 element of the estimate error covariance matrix in Equation~(\ref{Sigma})] of 0.2.
Fig.~\ref{fig_lincoef} shows the estimates from one of the runs and the corresponding true values of the linear coefficients $\{a_k\}$.  
The error bars, which are smaller than the size of data points, come from the square-root of the diagonal of the estimator covariance matrix.
Similarly, Fig.~\ref{fig_quadcoef} shows the estimates and true values of the quadratic coefficients $\{a_ka^\star_j\}$ (although complex conjugation was not needed since all coefficients were real).

Experiment 2 was the same as experiment 1, except with only 1000 exposures, with a mean Strehl ratio of 0.76, were used.  
Since Experiment 2 used one-quarter of the exposures used in Experiment 1, a simple $1/T$\ dependence of the estimator variance would lead to standard deviations that are double those of Experiment 1.
Experiment 2 was run 20 times, each with a different realization of the random noise described above.
The mean estimate of the planetary intensity in these runs was 0.7 photons/ms and a standard deviation was 0.96, comparable to the analytical result of 1.1.
Comparing the analytical standard deviations from Experiments 1 and 2, one can see that $1.1 >> 2 \times 0.2$, supporting the contention in Equation~(\ref{average}) that the variance of the estimate decreases more quickly than $1/T$.
The estimates of the static aberration coefficients had analytical standard deviations about 6 times those in Experiment 1.  The equivalent to Fig.~\ref{fig_lincoef} is visually undistinguishable, as the error  bars are still quite small.
The average analytical variance of the 121 quadratic aberration coefficients was about 2.7 times that of those in Experiment 1, again supporting Equation~(\ref{average}).

Experiment 3 was designed to test the algorithm's stability to planetary emission from angles other than $\balpha_p$, to help evaluate its utility as a scanning detector, in which one moves the scanning template and reprocesses the data for each template location.
Ideally, planetary emission from outside the scanning template would have a negligible effect on the estimates of both the static aberration coefficients and the planetary emission at $\balpha_p$.
Experiment 3 was the same as Experiment 2, except the planet was placed at an angular distance of about $1.75 \lambda/D$\ (7 pixels on the image grid) to the right of the assumed planetary position (i.e., that of Experiments 1 and 2).
The same matrix $\bH$\ from Experiment 2 was used, which assumed any planetary emission was coming from the planet location in Experiments 1 and 2 (see Figure~\ref{fig_ftqsa}).
Thus, the matrix $\bH$\ assumed that the planet was in the wrong location. 
In addition, the planet's brightness was increased to 1000 photons/ms, reducing the contrast ratio to $10^2$, and making the planet easily visible in the time-averaged image.
The ideal result of this experiment would be a planetary intensity estimate of 0 and the correct estimates of the static aberration coefficients.
The estimated planetary intensity was about 25 photons/ms, or 2.5\% of the true planetary brightness.  
Thus, this experiment showed 2.5\% leakage from a distance of $1.75 \lambda/D$, and the scanning detector response is strongly peaked at the correct location.   
Importantly, in this experiment, the estimates of the aberration coefficients were nearly as good as in Experiment 2.   
When the planet was placed at the smaller distance of about $0.9 \lambda/D$, the leakage was much greater - about 12\% and the static aberration coefficients where not as accurate.
This indicates that scanning template of a size of several $\lambda/D$\ should be sufficient to implement a scanning detector/estimator.
Such a scanning detector could be run iteratively, with the second iteration using a spatial template that includes all likely source locations from the first.

\section{Conclusions}\label{Conclusions}

\begin{figure}
\epsfig{file=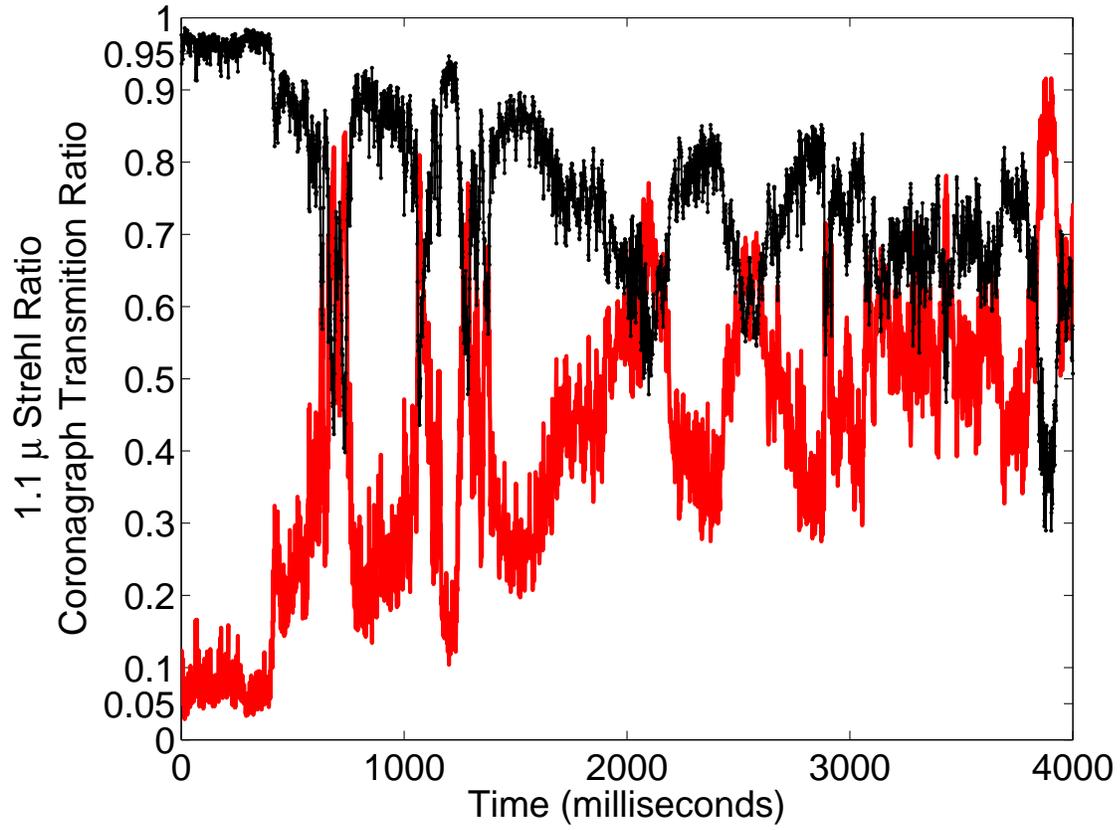,width=.9\linewidth,clip=}
\caption{\small (color online) The $2.2 \, \mu$\ The black curve shows Strehl ratio for the Experiment 1, and the red curve below shows the fraction of light removed by the ideal coronagraph.}
\label{fig_strehl}
\end{figure}

The purpose of this paper is to illustrate the possible potential for short-exposure imaging of exoplanets and thereby spur further research on the topic of short exposure imaging for high contrast applications.
Specifically, it describes a method for simultaneously estimating the static aberrations and the planetary emission from millisecond exposures with  ground-based stellar coronagraphs.
The ability to estimate planetary intensities in as little as few seconds has the potential to greatly improve the efficiency of exoplanet search surveys.
The new and crucial aspect of this work is that the method incorporates the wavefront sensor's estimate of the AO residual phase to provide much needed diversity in the data.
Thus, new and statistically independent information about the planetary emissions and quasi-static aberrations is revealed on the atmospheric clearing time-scale (McIntosh et al.~2005), which is, more-or-less, the telescope diameter divided by the wind speed.   
It is this diversity property that allows the covariance of the estimate of the planetary emission and the QS aberrations to decrease more quickly than inversely proportionally to the observation time.
While the random AO residual is an extremely useful source of diversity that can be used to characterize the optical system while simultaneously estimating the planetary emission, a fundamental trade-off is shown in Fig.~\ref{fig_strehl}: Large AO residual is accompanied by poor rejection of stellar emission by the coronagraph.
 
Due to the exploratory and preliminary nature of this work, many assumptions were made for the sake of simplicity. 
Among these assumptions are the the lack of aberrations downstream of the coronagraph, an ideal coronagraph, Fraunhofer diffraction, and no temporal variation in the static aberrations, to name a few.
The model can be generalized to incorporate more effects.
For example, it can allow for temporal variation of the quasi-static aberrations on longer time-scales using a scheme such spline interpolation in the time dimension, at the expense of having to estimate the spline coefficients.  
Furthermore, the method given here generalizes to the include spectrally resolved data and the diurnal field rotation.
E.g., to take advantage of $S$\ spectral channels, one must solve for $S$\ planetary intensities simultaneously, but there are still only $K$\ independent values characterizing the QS aberrations, since the phase variation scales with wavelength.
Similarly, to take advantage of diurnal field rotation (angular diversity), one only has to apply a coordinate transformation before including the additional data.

The fundamental limits of the short exposure approach are not easy to determine, as they are set by the accuracy of the various modeling aspects of the coronagraph system, including the reduction of the QS aberrations to a finite set of parameters,  as well as the various/bias characteristics of the estimator of the AO residual.   In a practical application, these will need to be evaluated with care.

In order to apply the method proposed here to real data sets a number of challenges must be met.   While any practical realization will no doubt have to consider various details, e.g., time differences in the exposures of the WFS and the science camera, the more essential tasks include the following:
\begin{itemize}
\item{Develop detectors with low readout noise so that short exposures are not penalized heavily.  Shorter wavelengths at which readout noise is less of a challenge should also be considered.}
\item{Provide an expansion of the QS aberrations that is realistic, as per Eq.~(\ref{phi_expansion}).
As there is no requirement that the expansion functions are orthogonal, there is considerable freedom allowed.
However, the more terms that are required, the more difficult the estimation task becomes.  
Any method for estimation of exoplanet intensities, whether with long exposures or short, must do this in some form.}
\item{Find  numerical methods for handling the quadratic system in Section~\ref{quadratic}.
The number of unknowns in the linear system in Section~\ref{linear} grows quadratically with the number of unknown aberration coefficients, while in the quadratic formulation, it grows only linearly.}
\item{Provide optimal estimators of AO residual $\phi_r(\brho,t)$ and its error statistics, given the WFS data.}
\item{Determine the bias and variance of the elements of the matrix $\bH$\ and the atmospheric speckle function $\mathcal{A}(\brho,t)$,
given the statistics of the AO residual.}
\item{Determine the error bars on the estimated quantities within the linear and quadratic formulations, given the required statistics on the necessary quantities.}
\item{In order to reduce memory requirements and processing times, sequential estimation methods in the spirit of Kalman filters should be considered. }
\end{itemize}

I would like to thank the anonymous referee for greatly improving the quality of this paper, and  Paul Shearer, Yves Atchade, Jr., Peter Lawson, Jean-Fran\c{c}ois Sauvage, Kjethil Dohlen, and Mamadou Ndiaye for discussions that enriched these investigations.
I extend special gratitude to Lewis Roberts for help with the AEOS data. 
This research was partially supported by a grant to the University of Michigan from the National Academies Keck Futures Initiative Spring 2011.
 
\clearpage

\clearpage


Barrett, Harrison H, Dainty, Christopher, Lara, David, ``Maximum-likelihood methods in wavefront sensing: stochastic models and likelihood functions," JOSA A, Vol. 24 Issue 2, pp.391-414 (2007)

B\'echet, Cl\'ementine; Tallon, Michel; Thi\'ebaut, \'Eric, ``Comparison of minimum-norm maximum likelihood and maximum a posteriori wavefront reconstructions for large adaptive optics systems," JOSA A, Vol. 26 Issue 3, pp.497-508 (2009)

Bloemhof, E.E., ``Anomolous intensity of spinned speckles at high adaptive correction," Optics Letters, vol. 29, Issue 2, pp. 159-161 (2004)


Boccaletti, A., Labeyrie, A. \& Ragazzoni, R. 1998, Astron. Astrophys., 338, 106

Caucci, L., Barrett, H.H., Devaney, N., Rodr\'iguez, J.J., ``Application of the Hotelling and ideal observers to detection and localization of exoplanets,'' JOSA A, Vol. 24, Issue 12, pp. B13-24 (2007)

Caucci, L., Barrett, H.H., Rodr\'iguez, J.J., ``Spatio-temporal Hotelling observer for signal detection from image sequences,'' Optics Express, Vol. 17, Issue 13, pp. 10946-10958  (2009)

Cavarroc, C., Boccaletti, A., Baudoz, P., Fusco, T. \& Rouan, D. 2006, Astron. Astrophys., 447, 397


Codona, J.L., Kenworthy, M.A., Lloyd-Hart, M. 2008, Proc. SPIE, 7015, 70155D


Gladysz, S., Yaitskova, N., Christou, J.C., ``Statistics of intensity in adaptive-optics images and their usefulness for detection and photometry of exoplanets," JOSA A, Vol. 27, Issue 11, pp.64-75 (2010)

Goodman, J.W., Introduction of Fourier Optics (New York: McGraw-Hill), 1968.

Goodman, J.W., Statistical Optics (New York: John Wiley \& Sons, Inc.), 1985, p. 466.

Goodman, J.W., Speckle Phenomena in Optics: Theory and Applications (Englewood, CO: Roberts \& Co.), 2007, p. 28.

Labeyrie, A., ``Images of exo-planets obtainable from dark speckles in adaptive optics,'' Astronomy \& Astrophysics, Vol. 298, pp. 544-548 (1995)

Lafreni\`ere, D., Marois, C., Doyon, R., Nadeau, D., Artigau, \'E., ``A New Algorithm for Point-Spread Function Subtraction in High-Contrast Imaging: A Demonstration with Angular Differential Imaging,'' ApJ, 660, 770 (2007)

Lyot, B., ``The study of the solar corona and prominences without eclipses," MNRAS, Vol. 99, pp. 580-594 (1939)

Marois, C., Lafreni\`ere, D., Doyon, R., Macintosh, B., Nadeau, D., ``Angular Differential Imaging: A Powerful High-Contrast Imaging Technique,'' ApJ, 641, 556 (2006)

Roberts, L.C., Jr., Neyman, C., ``Characterization of the AEOS Adaptive Optics System," PASP, vol. 114, pp.1260-1266 (2002)

Roggemann, M.C. and Meinhardt, J.A., ``Image Reconsruction by means of wave-front sensor measurements in closed-loop adaptive-optics systems," JOSA A vol. 10, pp. 1996-2007 (1993)

Sauvage, J.-F., Mugnier, L.M., Rousset, G., Fusco, T., ``Analytical expression of long-exposure adaptive-optics-corrected coronagraphic image. First application to exoplanet detection," JOSA A, Vol. 27, Issue 11, pp. 157-170 (2010)

Soummer, R., Pueyo, L., Larkin, J., ``Detection and Characterization of Exoplanets and Disks using Projections on Karhunen-Lo`eve Eigenimages," ApJL, {\it In Press} (2012)

Tropp, J.A. 2012, ``User-Friendly Tail Bounds for Sums of Random Matrices,'' Foundations of Computational Mathematics, 12, 389

Unser, M., ``Splines: a perfect fit for signal and image processing," IEEE Signal Processing Magazine, 16, 22-38 (1999)

Van Huffel, Sabine, and Vanderwalle, Joos, eds., ``The Total Least Squares Problem: Computational Aspects and Analysis" (SIAM, Philadelphia, 1991).

\end{document}